\newcounter{myctr}
\def\myitem{\refstepcounter{myctr}\bibfont\noindent\ifnum\themyctr>9\else\phantom{0}\fi\hangindent17pt\themyctr.\enskip}

\documentclass{ws-ijqi}

\usepackage{color}
\usepackage{float}
\usepackage{amsmath}
\usepackage{amssymb}
\usepackage{amsfonts}
\usepackage{latexsym}
\usepackage{mathrsfs}
\usepackage{bm}

\newcommand{\be}{\begin{eqnarray}}
\newcommand{\ee}{\end{eqnarray}}
\newcommand{\bea}{\begin{eqnarray*}}
\newcommand{\eea}{\end{eqnarray*}}
\def\({\left(}
\def\){\right)}
\def\[{\left[}
\def\]{\right]}
\def\C{\mathbb{C}}

\def\Z{\mathbb{Z}}
\newcommand{\im}{\mathrm{i}}

\newcommand{\braket}[1]{\left\langle #1 \right\rangle}

\newcommand{\bra}[1]{\langle #1 |}
\newcommand{\ket}[1]{| #1 \rangle}

\newcommand{\sla}[1]{\rlap{\kern .15em /}#1}

\newcommand{\ot}{\otimes}
\newcommand{\Bot}{\bigotimes}

\def\fp#1#2#3{{\it Fortschr.\ Phys.}\ {\bf {#1}} ({#2}) #3}

\def\prA#1#2#3{{\it Phys.\ Rev.}\ {\bf A{#1}} ({#2}) #3}

\def\prl#1#2#3{{\it Phys.\ Rev.\ Lett.}\ {\bf #1} ({#2}) #3}

\begin{document}

\markboth{Tanaka {\it et al.}}
{Quantum Oracles in Terms of Universal Gate Set}

\catchline{}{}{}{}{}

\title{QUANTUM ORACLES IN TERMS OF UNIVERSAL GATE SET}

\author{YUJI TANAKA\footnote{Present address: Department of Physics, Graduate School of Science, Osaka University, 1-1 Machikaneyama, Toyonaka, Osaka 560-0043, Japan.}}

\address{Department of Physics, Kinki University, 3-4-1 Kowakae, Higashi-Osaka, Osaka 577-8502, Japan\\
tanaka@cmp.sanken.osaka-u.ac.jp}

\author{TSUBASA ICHIKAWA}

\address{Research Center for Quantum Computing, Interdisciplinary Graduate School of Science and Engineering, Kinki University, 3-4-1 Kowakae, Higashi-Osaka, Osaka 577-8502, Japan}

\author{MASAHITO TADA-UMEZAKI\footnote{Present address: Division of International Cooperative Research,
Research Center for Ethnomedicine,
Institute of Natural Medicine,
University of Toyama,
2630 Sugitani, Toyama, 930-0194, Japan.}}

\address{Research Center for Quantum Computing, Interdisciplinary Graduate School of Science and Engineering, Kinki University, 3-4-1 Kowakae, Higashi-Osaka, Osaka 577-8502, Japan}

\author{YUKIHIRO OTA}

\address{CCSE, Japan Atomic Energy Agency, 6-9-3 Higashi-Ueno Taito-ku, Tokyo 110-0015, Japan\\
 and CREST(JST), 4-1-8 Honcho, Kawaguchi, Saitama 332-0012, Japan}

\author{MIKIO NAKAHARA}

\address{Department of Physics, Kinki University, 3-4-1 Kowakae, Higashi-Osaka, Osaka 577-8502, Japan\\
 and Research Center for Quantum Computing, Interdisciplinary Graduate School of Science and Engineering, Kinki University, 3-4-1 Kowakae, Higashi-Osaka, Osaka 577-8502, Japan}

\maketitle

\begin{history}
\received{Day Month Year}
\revised{Day Month Year}
\end{history}

\begin{abstract}
We present a systematic construction of quantum circuits implementing Grover\rq s database search algorithm for arbitrary number of targets. We introduce a
new operator which flips the sign of the targets and evaluate its circuit complexity. We find the condition under which the circuit complexity of the database search algorithm based on this operator is less than that of the conventional 
one.
\end{abstract}


\section{Introduction}
\setcounter{equation}{0}
Efficiency of quantum algorithm, {\it e.g.} speed-up, over its classical counterpart can be found typically in two algorithms: factorization of large numbers and database search. Shor\rq s algorithm$^1$  solves the former problem with exponential efficiency compared to its classical counterpart, while Grover's algorithm (GA)$^{2 \text{--} 7}$ solves the latter with quadratic efficiency. GA has been a fascinating one to be implemented, since a lot of informational problems are translated into the database search problem, in spite of its moderate speed-up.

There are two different classes in the database search problem. One is a purely information theoretic problem such as satisfiability problem, where a Boolean function $f$ is given and bit strings satisfying $f$ should be identified. The other is a rather practical unstructured database search problem, {\it e.g.} 
to find the owner of a telephone in a telephone directory, given a phone number, as stated in Ref.~2. It should be stressed that, in 
contrast to the former, the player is allowed to know what should be searched
(the owner of a telephone) in advance in the latter case and 
searches the target under the reference of this information.
We focus on the latter situation in this paper, 
motivated by the fact that  many experimental demonstrations and proposals of implementation of GA$^{7\text{--}16}$ implicitly assume the latter situation.
Also, these implementations so far focus only on GA with a single target and there seems no systematic construction of a quantum circuit implementing GA with multiple targets. The detailed analysis of 
the circuit complexity has not been presented for this case.
 
In this paper, by a simple way, we provide a systematic construction of quantum circuits which implements GA with multiple targets. First, we propose a unitary operator which replaces the conventional oracle operator of GA (Theorem 1).  We show that the construction of a quantum circuit of our oracle operator amounts to implementing a unitary operator which yields an equal weight superposition state of the target states from $\ket{0}^{\otimes n}$. Utilizing the dichotomy on each bit in the target sequences, we obtain a simple recursion relation with which the quantum circuit of the unitary operator is designed (Lemma 2 and Theorem 2). Also, we propose another database search method (Theorem 3), in which circuit complexity is reduced considerably (Proposition 2). Hence, the present systematic and explicit \lq algorithm\rq~to build the oracle in the quantum search enables us to estimate the circuit complexity.

This paper is organized as follows. In Sec.~2, we show the equivalence between the proposed oracle operator and the conventional one. 
Quantum circuit design of the former one and its analysis from the viewpoint of the circuit complexity are provided in Sec.~3 and possibility of further reduction of the gate complexity is discussed in Sec.~4. Simple examples are given in Sec.~5. Section~6 is devoted to conclusion and discussions.

\section{Equivalent Quantum Algorithms}
Consider a database  which consists of $2^n$ elements each of which is labeled by
\be
x:=\sum_{i=0}^{n-1}x_i2^{i},
\label{binary}
\ee
where $x_i \in \{0,1\}$ for all $i$. The set of all labels is denoted by $N:=\{0,1,\ldots, 2^n-1\}$. A nonempty subset $S$ of $N$ is made of the set of the targets of the database search problem. We further introduce the complementary subset $\bar{S}:=N\backslash S$.
Utilizing the binary representation (\ref{binary}) of $x$, we can represent each element $x\in N$ as an $n$-qubit normalized quantum state in ${\cal H}:=\bigotimes_{i=1}^n\mathcal{H}_i$ with $\mathcal{H}_i=\C^2$ for all $i$,
\be
\ket{x}:=\ket{x_{n-1}}_1\ot\ket{x_{n-2}}_2\ot\cdots\ot\ket{x_0}_n,
\,\,
\ket{x_{n-i}}_i\in\mathcal{H}_i,
\label{xstate}
\ee
where
\bea
{}_{n-i}\braket{x_{i}|y_{i}}{}_{n-i}=\delta_{x_{i}y_{i}},
\qquad
x_{i},y_{i}\in \{0,1\}
\eea
for all $i$ and $\ket{x_i}_{n-i}\in{\cal H}_{n-i}$ is a normalized eigenvector of the Pauli matrix $\sigma_z$ with the eigenvalue $(-1)^{x_i}$. Note that the set $\{\ket{x}\,|\,x\in N\}$ constitutes a complete orthonormal basis in ${\cal H}$. Let
\be
\ket{\psi}:=|N|^{-1/2}\sum_{x\in N}\ket{x},
\label{psi}
\ee
be the uniform superposition of all the basis vectors of $\mathcal{H}$,
where $|A|$ denotes the cardinality of a set $A$. The state $\ket{\psi}$ is the initial state in GA. 

\begin{figure}[t]
\begin{center}
\includegraphics[width=3in]{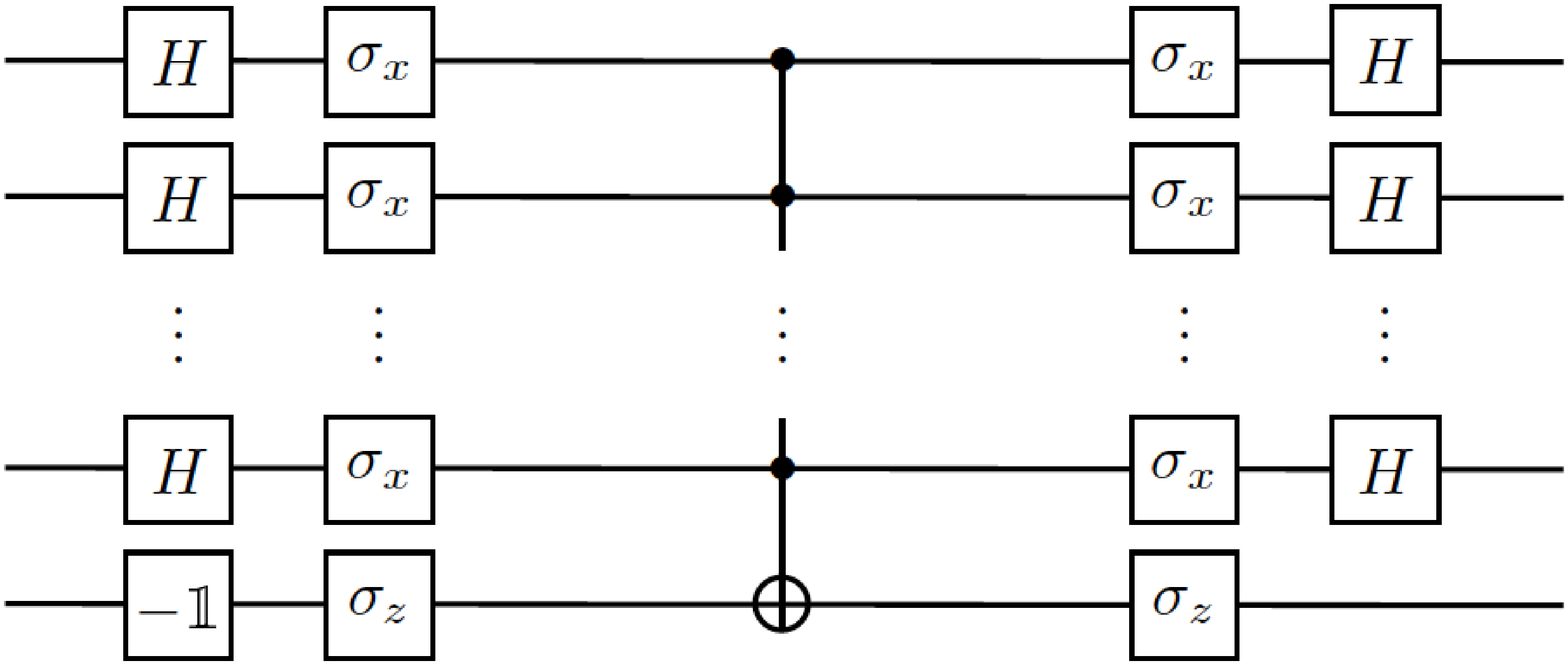}
\caption{Quantum circuit for the inversion-about-mean operator $D$. Gate $H$ denotes Hadamard transformation given in Eq.~(\ref{Hadamard}).}
\label{D}
\end{center}
\end{figure}

Let us define two unitary operators essential for GA,
 the inversion operator $D$ with respect to the mean and the conventional 
oracle operator $O_{\rm conv}(S)$, respectively, as
\be
D:=-\bm{1}^{\ot n}+2\ket{\psi}\bra{\psi},
\label{defD}
\ee
and
\bea
O_{\rm conv}(S):=\bm{1}^{\ot n}-2\sum_{x\in{S}}\ket{x}\bra{x},
\eea
where
$\bm{1}$ is the identity operator on $\C^2$. Note that the oracle operator $O_{\rm conv}(S)$ distinguishes the targets in the set $N$ as
\bea
O_{\rm conv}(S)\ket{x}=
\begin{cases}
-\ket{x}&x\in S, \\
\ket{x}&x\in\bar{S}\\
\end{cases}.
\eea
Then, measurements of the state 
\be
\ket{\psi^{(k)}}:=[DO_{\rm conv}(S)]^k\ket{\psi}
\label{ConvGrover}
\ee
with $k\approx{\cal O}(\sqrt{|N|/|S|})$ yield elements of $S$ with high 
probability$^{3\text{--}7}$. Figure \ref{D} shows a
quantum circuit which implements $D$.~{$^{7,10}$}

Although the oracle operator $O_{\rm conv}(S)$ has been utilized as a {\it de facto} standard, we introduce another oracle operator. Let $\ket{S}$ be
\be
\ket{S}:=|S|^{-1/2}\sum_{x\in S}\ket{x}
\label{stateS}
\ee
and define
\be
O(S):=\bm{1}^{\ot n}-2\ket{S}\bra{S},
\label{defOS}
\ee
which distinguishes the targets from the others and plays the central r\^ole 
in the present work. The operator
$O(S)$ replaces $O_{\rm conv}(S)$ as we show next.

Let us first show that the action of $O(S)$ is equivalent to that of $O_{\rm conv}(S)$ on the specific state $\ket{\psi}$ in Eq.~(\ref{psi}). To this end,
we prove
\begin{lemma}
The state $\ket{\phi}\in{\cal H}$ is an element of the kernel of $O(S)-O_{\rm conv}(S)$ if and only if $\ket{\phi}$ can be written as
\bea
\ket{\phi}=c\ket{S}+\sum_{x\in \bar{S}}c_x\ket{x},
\eea
which implies
\bea
{\rm Ker}[O(S)-O_{\rm conv}(S)]=\C^{|\bar{S}|+1}.
\eea
\end{lemma}
\begin{proof}
Let
\bea
\ket{\phi}=\sum_{x\in N}c_x\ket{x}
\eea
be any vector in $\mathcal{H}$.
Then, we find
\bea
O(S)\ket{\phi}=\ket{\phi}-2|S|^{-1/2}\sum_{x\in S}c_x\ket{S}
\eea
and
\bea
O_{\rm conv}(S)\ket{\phi}=\ket{\phi}-2\sum_{x\in S}c_x\ket{x}.
\eea
Let $\ket{\phi}$ be an element of the kernel of $O(S)-O_{\rm conv}(S)$,
that is,
\bea
\bra{y}[O(S)-O_{\rm conv}(S)]\ket{\phi}=0
\eea
for all $y\in N$. Then we obtain for any $z\in S$
\bea
\sum_{x\in S}c_x=|S|c_z,
\eea
from which we find
\bea
\ket{\phi}\in{\rm Ker}[O(S)-O_{\rm conv}(S)]\,\,\Rightarrow\,\, \ket{\phi}=c\ket{S}+\sum_{x\in \bar{S}}c_x\ket{x},
\eea
where $c$ is the common coefficient of $\ket{x}, x \in S$, in $\ket{\phi}$.
The converse is trivial.
\end{proof}

\smallskip

At this stage, we note that, introducing real parameters $\varphi_k$ and 
the state $\ket{\bar{S}}$ constructed by the same fashion as in 
Eq.~(\ref{stateS}), 
the state $\ket{\psi^{(k)}}$ can be rewritten as
\bea
\ket{\psi^{(k)}}=\sin\varphi_k\ket{S}+\cos\varphi_k\ket{\bar{S}}
\eea
as shown in Ref.~4. Thus, we immediately find
\begin{corollary}
The actions of $O(S)$ and $O_{\rm conv}(S)$ on $\ket{\psi^{(k)}}$ result in the same state:
\bea
O(S)\ket{\psi^{(k)}}=O_{\rm conv}(S)\ket{\psi^{(k)}}
\eea
for any positive integer $k$.
\end{corollary}

\smallskip

Since this Corollary is valid for all $k \in \mathbb{N}$, we may replace all actions of $O_{\rm conv}(S)$ by $O(S)$ during the iteration processes in GA. 
Hence, we conclude that 
\begin{theorem}
Grover\rq s algorithm can be realized by the iterations of $DO(S)$, that is,
\bea
\ket{\psi^{(k)}}=[DO(S)]^k\ket{\psi}.
\eea
\end{theorem}
\smallskip

This theorem suggests non-uniqueness of the way to realize quantum information processing. 

\section{Building Quantum Circuits by Dichotomy}

We turn into an explicit and systematic construction of the oracle $O(S)$ via elementary quantum gates. The key observation is that $O(S)$ is unitarily 
equivalent to a conventional oracle operator in GA with 
a unique target $\ket{0}^{\ot n}\in{\cal H}$. Let
\bea
P:=\bm{1}^{\ot n}-2(\ket{0}\bra{0})^{\ot n}
\eea
be this conventional oracle operator
and define $U(S)$ by 
\be
\ket{S}=U(S)\ket{0}^{\ot n}.
\label{USn}
\ee
Then we find that $O(S)$ and $P$ are related by $U(S)$ as
\be
O(S)=U(S)PU(S)^\dag,
\label{decO}
\ee
which can be interpreted as a decomposition of $O(S)$ in terms of $U(S)$ and $P$.

\begin{figure}[t]
\begin{center}
\includegraphics[width=3in]{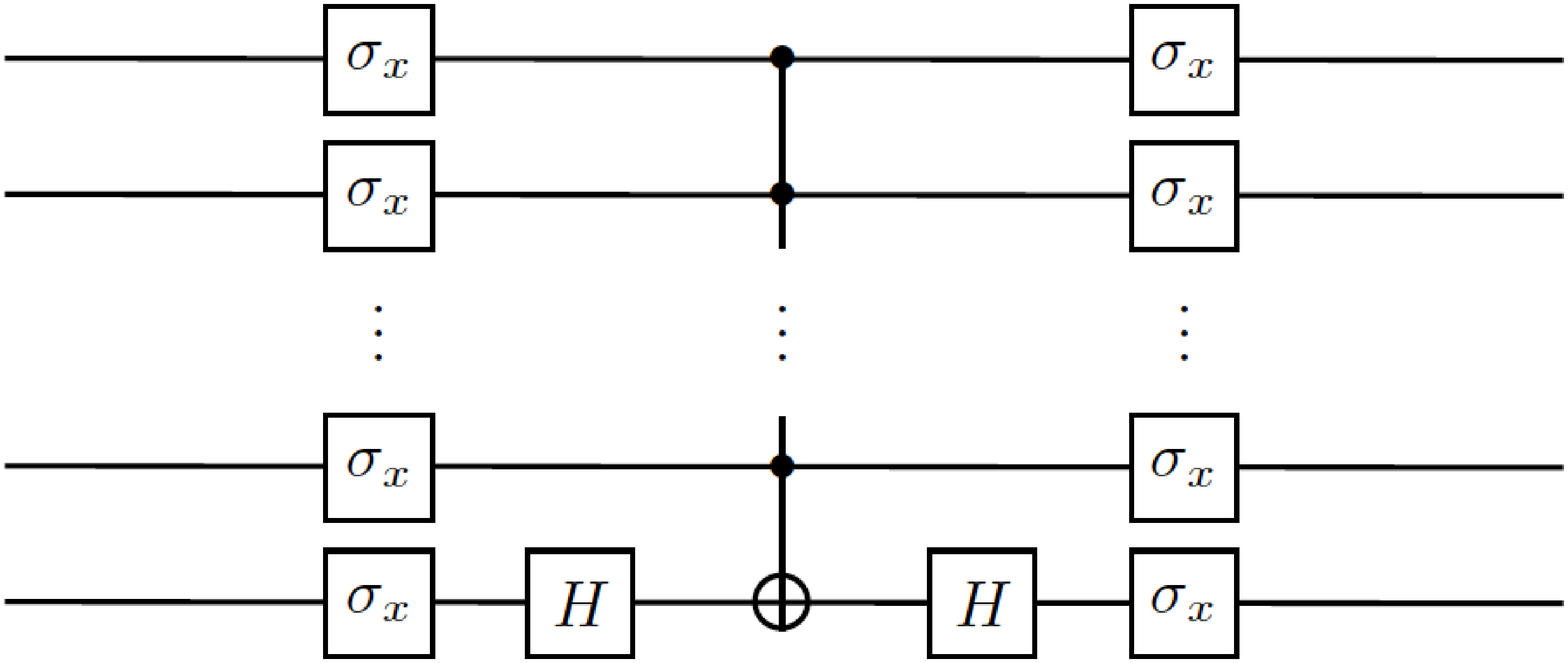}
\caption{Quantum circuit for $P$.}
\label{P}
\end{center}
\end{figure}

We address two features of the decomposition (\ref{decO}). First, the unitary
operator $U(S)$ in Eq.~(\ref{USn}) is not uniquely determined since it does
not specify how basis vectors other than $\ket{0}^{\otimes n}$
are mapped under $U(S)$.
Thus, we define a set ${\cal S}(S)$ whose elements are $U(S)$: 
\bea
{\cal S}(S):=\{U(S)\,|\, \ket{S}=U(S)\ket{0}^{\ot n}\} \simeq {\mathrm{U}}(2^n-1),
\eea
which has $(2^n-1)^2$ free parameters.
Second, since the construction of $D$ and $P$ from the universal gate set 
(local unitary operators and CNOT gates) has been clarified in Refs.~7 and 10
(See Fig. \ref{P}), the construction of quantum circuits for GA boils down to 
that of $U(S)$. Note that from Fig.~\ref{P} and Ref.~17, the circuit
complexity for $P$ is at most ${\cal O}(n^2)$. Utilizing the operator $P$, 
we also find a circuit diagram of the conventional
oracle $O_{\rm conv}(S)$ with the 
universal gate set as a biproduct, which is given in Appendix.

To proceed, let us classify elements of $S$ by dichotomy on the value of each bit. Prior to this, 
let us define, for a given $x \in N$,
\bea
x^{(m)}:=\sum_{i=1}^mx_{n-i}2^{m-i}
\eea
for an integer $m=1,2,\dots,n$, which can be embedded in
$\Bot_{i=1}^m{\cal H}_i$ as a normalized vector
\bea
\ket{x^{(m)}}:=\ket{x_{n-1}}_1\ot\ket{x_{n-2}}_2\ot\cdots\ot\ket{x_{n-m}}_m.
\eea
This vector $\ket{x^{(m)}}$ is constructed by picking up the first $m$ bits 
of the binary representation of $x$ and utilizing the correspondence similar 
to Eq.~(\ref{xstate}). Note that $x^{(n)}=x$.
We introduce two types of subsets of $S$ with a help of $x^{(m)}$: 
One is
\be
S_m(\alpha_m):=\{x\,|\, x^{(m)}=\alpha_m, x\in S\}
\label{defSalpha}
\ee
for $m=1,2,\dots, n$ and $\alpha_m=0,1,\dots,2^{m}-1$. The other results from 
the dichotomy:
\bea
S_{m+1}(i, \alpha_m)
:=\{x\,|\, x_{n-m-1}=i, x\in S_m(\alpha_m)\},
\eea
for $i=0,1$ and $m=1,2,\dots, n-1$. 

Next, we focus on relations between the cardinalities of the above two sets $S_m(\alpha_m)$ and $S_{m+1}(i,\alpha_m)$. The following three identities are 
easy to verify:
\begin{subequations}
\begin{eqnarray}
&&|S_m(0)|+|S_m(1)|+\cdots+|S_m(2^m-1)|=|S|, \label{idone}\\
&&|S_{m+1}(0, \alpha_m)|+|S_{m+1}(1, \alpha_m)|=|S_m(\alpha_m)|,\label{idtwo}\\
&&|S_{m+1}(i, \alpha_m)|=|S_{m+1}(2 \alpha_m+i)|. \label{idthree}
\end{eqnarray}
\end{subequations}
We can formally introduce a ``probability'' distribution $\{p_m(\alpha_m)\}_{\alpha_m=0}^{2^m-1}$
\bea
p_m(\alpha_m):=\frac{|S_m(\alpha_m)|}{|S|},
\qquad
\alpha_m=0,1,\dots,2^m-1,
\eea
from Eq.~(\ref{idone}), and a conditional probability distribution $\{p_{m+1}(i|\alpha_m)\}_{i=0,1}$
\bea
p_{m+1}(i|\alpha_m):=\frac{|S_{m+1}(i, \alpha_m)|}{|S_m(\alpha_m)|},
\qquad
i=0,1
\eea
from Eq.~(\ref{idtwo}). 
Note that Eq.~(\ref{idthree}) implies 
\be
p_{m+1}(i|\alpha_m)p_m(\alpha_m)=p_{m+1}(2\alpha_m+i).
\label{Bayes}
\ee
Let $A_m$ be a set satisfying
\bea
A_m:=\{\alpha_m\,|\,S_m(\alpha_m)\neq\emptyset\}.
\eea
Then, we have
\be
p_m(\alpha_m)=0
\quad
{\rm for}
\quad
\alpha_m\not\in A_m.
\label{probA}
\ee

We show a recursion relation to construct $U(S)$. First, on the basis of the above probability distributions, we define two kinds of unitary operators on $\C^2$
by
\be
V_1:=\sqrt{p_1(0)}\bm{1}-\im\sqrt{p_1(1)}\sigma_y
\label{V1}
\ee
and
\be
V_{m+1}(\alpha_m)=\sqrt{p_{m+1}(0|\alpha_m)}\bm{1}-\im\sqrt{p_{m+1}(1|\alpha_m)}\sigma_y,
\label{Vm}
\ee
for $\alpha_m\in A_m$, whereas we define $V_{m+1}(\alpha_m)=\bm{1}$ for 
$\alpha_m\not\in A_m$.
These one qubit unitary operators are followed by unitary operators
\bea
U_1:=V_1\ot\bm{1}^{\ot(n-1)}
\eea
and
\be
U_{m+1}(\alpha_m)&:=\ket{\alpha_m}\bra{\alpha_m}\ot V_{m+1}(\alpha_m)\ot\bm{1}^{\ot(n-m-1)}\nonumber\\
&+\(\bm{1}^{\ot m}-\ket{\alpha_m}\bra{\alpha_m}\)\ot\bm{1}^{\ot(n-m)},
\label{Ualpha}
\ee
respectively. Here, $\ket{\alpha_m}$ is a normalized vector in $\Bot_{i=1}^m{\cal H}_i$ constructed from $\alpha_m$ by the same way as $\ket{x^{(m)}}$.  
Note that the elements of the set $\{U_{m+1}(\alpha_m)\}_{\alpha_m}$ are 
commutative with each other:
\bea
[U_{m+1}(\alpha_m), U_{m+1}(\beta_m)]=0.
\eea
Following this, let us define
\be
U_{m+1}:=\prod_{\alpha_m=0}^{2^m-1}U_{m+1}(\alpha_m)=\sum_{\alpha_m=0}^{2^m-1}\ket{\alpha_m}\bra{\alpha_m}\ot V_{m+1}(\alpha_{m})\ot\bm{1}^{\ot(n-m-1)}
\label{Um}
\ee
for $m=1,2,\dots, n-1$. Operators $U_{m+1}$ are nothing but
uniformly controlled gates introduced in Refs. 18 and 19. Then, we have
\begin{lemma}
\be
U_{m+1}U_{m}\cdots U_1\ket{0}^{\ot n}=\sum_{\alpha_{m+1}=0}^{2^{m+1}-1}\sqrt{p_{m+1}(\alpha_{m+1})}\ket{\alpha_{m+1}}\ot\ket{0}^{\ot (n-m-1)}
\label{Um_prod}
\ee
holds for $m=1, 2,\dots, n-1$.
\end{lemma}
\begin{proof}
We prove this claim by induction. For $m=1$, we can easily check the validity of Eq.~(\ref{Um_prod}).
Next, suppose that Eq.~(\ref{Um_prod}) holds for $m=k-1$. Then
using Eqs.~(\ref{Um}), (\ref{probA}), (\ref{Vm})  and (\ref{Bayes}), we obtain
\begin{eqnarray*}
U_{k+1}U_{k}\cdots U_1\ket{0}^{\ot n}&=&U_{k+1}\sum_{\alpha_{k}=0}^{2^{k}-1}\sqrt{p_{k}(\alpha_{k})}\ket{\alpha_{k}}\ot\ket{0}^{\ot (n-k)}\\
&=&\sum_{\alpha_{k}\in A_{k}}\sqrt{p_{k}(\alpha_{k})}\ket{\alpha_{k}}\ot V_{k+1}(\alpha_{k})\ket{0}\ot\ket{0}^{\ot (n-k-1)}\\
&=&\sum_{\alpha_{k}\in A_{k}}\sum_{i=0,1}\sqrt{p_{k+1}(i|\alpha_{k})p_{k}(\alpha_{k})}\ket{\alpha_{k}}\ot \ket{i}\ot\ket{0}^{\ot (n-k-1)}\\
&=&\sum_{\alpha_{k}\in A_{k}}\sum_{i=0,1}\sqrt{p_{k+1}(2\alpha_{k}+i)}\ket{\alpha_{k}}\ot \ket{i}\ot\ket{0}^{\ot (n-k-1)}.\\
\end{eqnarray*}
We find Eq.~(\ref{Um_prod}) with $m=k$ by introducing
$
\alpha_{k+1}=2\alpha_{k}+i$ and
$\ket{\alpha_{k+1}}=\ket{\alpha_{k}}\ot\ket{i}$,
since the probability distribution $p_{k+1}(2\alpha_{k}+i)$ is 
normalized by Eq.~(\ref{Bayes}).  This completes the proof.
\end{proof}

\smallskip

Let us construct
\bea
U:=U_nU_{n-1}\cdots U_1,
\eea
whose action on $\ket{0}^{\ot n}$ is found from Lemma 2 as
\bea
U\ket{0}^{\ot n}=\sum_{x\in N}\sqrt{p_n(x)}\ket{x}.
\eea
Since
\bea
p_n(x)=
\begin{cases}
0      & {\rm for}\,\, x\in \bar{S}, \\
|S|^{-1}      &{\rm for}\,\, x\in S
\end{cases},
\eea
is valid, we observe
\bea
U\ket{0}^{\ot n}=|S|^{-1/2}\sum_{x\in S}\ket{x}=\ket{S},
\eea
which manifestly proves
\begin{theorem}
The operator $U=U_nU_{n-1}\cdots U_1$ belongs to ${\cal S}(S)$, that is,
\bea
U\in{\cal S}(S).
\eea
\end{theorem}
\smallskip

\begin{figure}[t]
\begin{center}
\includegraphics[width=4in]{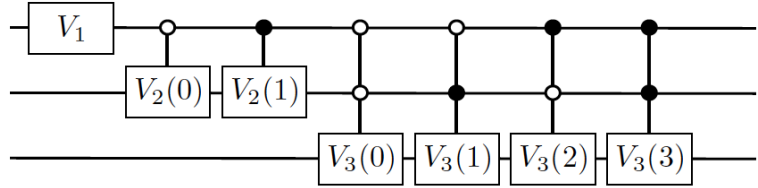}
\caption{Quantum circuit of the unitary operator $U(S)$ for $n=3$ in terms of
a one-qubit gate $V_1$ and controlled gates $U_{m+1}(\alpha_m)$. The one-qubit gates $V_{m+1}(\alpha_m)$ act on the $m+1$-th qubit if
the state of the control qubits has a non-vanishing support on $\ket{\alpha_m}$.}
\label{UTcircuit}
\end{center}
\end{figure}

Thus, hereafter we may put
\bea
U=U(S),
\eea
whose quantum circuit is given in Fig. \ref{UTcircuit}.

To close this section, let us estimate circuit complexity of $U(S)$, {\it i.e.}, the number of elements in the universal gate set necessary to emulate $U(S)$. This is done by noticing 
\be
A_n=S,
\label{AS}
\ee
which leads us to
\begin{proposition}
The circuit complexity of $U(S)$ has an upper bound ${\cal O}(n^3|S|/6)$.
\end{proposition}
\begin{proof}
Consider an element $\alpha_n\in A_n$. Then, there exists an appropriate doublet $(\alpha_{n-1}, i)$ satisfying 
\bea
\alpha_n=2\alpha_{n-1}+i.
\eea
Using Eqs.~(\ref{idtwo}) and (\ref{idthree}), we find that the existence of 
such a doublet implies $\alpha_{n-1}\in A_{n-1}$, which means that the 
operator (\ref{Vm}) is well-defined and might be non-trivial. Since this is true for all elements in $A_n$, from Eqs.~(\ref{Ualpha}) and (\ref{AS}), we obtain an upper bound $|S|$ for the number of non-trivial $U_n(\alpha_n)$. 
Also, by induction, we find that $|S|$ also becomes the upper bound of the
number of the non-trivial $U_m(\alpha_m)$ for any $m$.

Combining this bound $|S|$ with the fact that one-qubit unitary gates with $m$ control qubits can be emulated by ${\cal O}(m^2)$ elements of the universal gate set$^{17}$, we have an upper bound ${\cal O}(m^2|S|)$ on the gate complexity of $U_{m+1}$.
Summing up them from $m=1$ to $m=n-1$, 
we find the upper bound ${\cal O}(n^3|S|/6)$ for the circuit complexity of 
$U(S)$.
\end{proof}

\smallskip

We address that this bound is not so tight, since for the derivation of this 
bound we have not taken into account the possibilities that 
$p_{m+1}(1|\alpha_m)=0$ for $\alpha_m\in A_m$, which implies that 
$U_{m+1}(\alpha_m)$ becomes the identity operator.

\section{Reducing the Circuit Complexity}

Following the results in the previous section, let us present another 
construction of the quantum circuit and estimate its circuit complexity. For 
this purpose, we define a set $\tilde{S}$ by
\bea
\tilde{S}=\{0,1,\dots, |S|-1\}
\eea
which is related to the given $S$ by an appropriate element $\sigma$ of 
the symmetric group $\frak{S}_{|N|}$, that is,
\be
\sigma \tilde{S}=S.
\label{sigmaS}
\ee
Note that 
$
|\tilde{S}|=|S|.
$
We define $\ket{\tilde{S}}$ and $O(\tilde{S})$ by replacing $S$ with 
$\tilde{S}$ in Eqs.~(\ref{stateS}) and (\ref{defOS}), respectively. Also note 
that $\sigma$ satisfying Eq.~(\ref{sigmaS}) is not uniquely determined
since we do not specify how elements in $\bar{S}$ are mapped under $\sigma$.

Let $\pi_{\sigma}$ be an $|N|$-dimensional unitary representation of $\sigma$.
Then we observe
\begin{theorem}
Grover\rq s algorithm to amplify the target state $\ket{S}$ can be realized by the invocations of the oracle $O(\tilde{S})$, that is,
\be
 \ket{\psi^{(k)}}=\pi_\sigma[DO(\tilde{S})]^{k}\pi_\sigma^\dag\ket{\psi}.
 \label{NovelGrover}
 \ee
\end{theorem}
\begin{proof}
By the definition of $\tilde{S}$ and $\sigma$, there exists a
unitary operator $\pi_\sigma$ satisfying
\bea
\pi_\sigma U(\tilde{S})\in{\cal S}(S),
\eea
which implies
\bea
\pi_\sigma\ket{\tilde{S}}=\ket{S}
\qquad
{\rm and}
\qquad
\pi_\sigma O(\tilde{S})\pi_\sigma^\dag=O(S).
\eea
Since $\ket{\psi}$ in Eq.~(\ref{psi}) is manifestly a one-dimensional 
symmetric representation of $\frak{S}_{|N|}$, we obtain, by taking
Eq.~(\ref{defD}) into account, 
 \bea
 \pi_\tau\ket{\psi}=\ket{\psi}
 \eea
 and
 \bea
 [\pi_\tau,D]=0
 \eea
 for all $\tau\in\frak{S}_{|N|}$. Then, we observe
 \bea
 DO(S)=\pi_\sigma DO(\tilde{S})\pi_\sigma^\dag.
 \eea
 Using $\pi_\sigma^\dag\pi_\sigma=\bm{1}^{\ot n}$, we find
\begin{eqnarray*}
 \pi_\sigma[DO(\tilde{S})]^{k}\pi_\sigma^\dag\ket{\psi}
 &=[\pi_\sigma DO(\tilde{S})\pi_\sigma^\dag]^k\ket{\psi}
=[DO(S)]^{k}\ket{\psi}=\ket{\psi^{(k)}},
 \end{eqnarray*}
which completes the proof.
\end{proof}


\subsection{Construction of $O(\tilde{S})$}

Since one can learn from the above theorem that amplification of $\ket{S}$ can 
be performed by the invocations of the oracle $O(\tilde{S})$, let us 
investigate detailed properties of $O(\tilde{S})$ from the viewpoint of
circuit complexity. For this purpose, it is convenient to introduce a number 
$l\in\Z$ satisfying
\bea
l-1\le\log_2|\tilde{S}|< l,
\eea
which is the minimal number of bits required for binary representation
of all the elements of $\tilde{S}$. Thus, the $n$-bit representation (\ref{binary}) of each label of the element $x\in \tilde{S}$ should be
\bea
x_l=x_{l+1}=\dots=x_{n-1}=0.
\eea
Then, from Eq.~(\ref{defSalpha}), we find
\bea
S_m(\alpha_m)=
\begin{cases}
\tilde{S}      & {\rm for} \alpha_m=0, \\
\emptyset      & {\rm otherwise}
\end{cases},
\eea
for $m=1, 2, \dots, n-l+1$. Thus, if we employ the construction of $U(\tilde{S})$
introduced in the previous section, we observe
from Eqs.~(\ref{V1}) and (\ref{Vm}) that
\be
U_1=U_2=\dots=U_{n-l}=\bm{1}^{\ot n}
\label{redU}
\ee
for $\tilde{S}$. This implies that the circuit complexity of $U(\tilde{S})$ is considerably less than that of $U(S)$ for generic $S$.

Remarkably, there exists further reduction of the circuit complexity
as we will show now.
First, we fix notation. Associated with the subscript $m$ in 
Eq.~(\ref{defSalpha}), we employ a shifted index $m^{\prime}:=m-(n-l)$.
A normalized vector $\ket{\alpha_{m^\prime}}\in\Bot_{i=n-l+1}^m{\cal H}_i$ is defined by a similar manner to $\ket{\alpha_m}$, where $\alpha_{m^{\prime}} =0,1, \ldots, 2^{m^\prime}-1$. 

Next, for a given $\ket{\alpha_m}$ with $m\ge n-l$, we seek 
$\ket{\alpha_{m^{\prime}}}$ satisfying 
\be
\ket{\alpha_m}=\ket{0}_1\ot\ket{0}_2\ot\cdots\ot\ket{0}_{n-l}\ot\ket{\alpha_{m^\prime}}.
\label{aa}
\ee
Depending on a type of the solutions of Eq.~(\ref{aa}), we prepare a controlled unitary gate, $U^\prime_{m+1}(\alpha_m)$: (i) If there exists $\ket{\alpha_{m^\prime}}$ satisfying Eq.~(\ref{aa}) for $m^\prime >0$, then 
\begin{subequations}
\begin{eqnarray}
U^\prime_{m+1}(\alpha_m)&:=&\bm{1}^{\ot(n-l)}\ot\ket{\alpha_{m^\prime}}\bra{\alpha_{m^\prime}}\ot V_{m+1}(\alpha_m)\ot\bm{1}^{\ot (n-m-1)}\nonumber\\
&+&\bm{1}^{\ot(n-l)}\ot\(\bm{1}^{\ot m^\prime}-\ket{\alpha_{m^\prime}}\bra{\alpha_{m^\prime}}\)\ot\bm{1}^{\ot(n-m)}.
\label{Uprimealpha1}
\end{eqnarray}
(ii) If there exists $\ket{\alpha_{m^\prime}}$ satisfying Eq.~(\ref{aa}) for $m^\prime=0$, then 
\be
U^\prime_{m+1}(\alpha_m):=\bm{1}^{\ot(n-l)}\ot V_{m+1}(\alpha_m)\ot\bm{1}^{\ot(l-1)}.
\label{Uprimealpha2}
\ee
(iii) If there is no $\ket{\alpha_{m^\prime}}$ satisfying Eq.~(\ref{aa}), then 
\be
U^\prime_{m+1}(\alpha_m):=\bm{1}^{\ot n}.
\label{Uprimealpha3}
\ee
\end{subequations}
Under these preparations, we find
\begin{lemma}
The controlled gates $U_{m+1}(\alpha_m)$ for $\tilde{S}$ can be replaced with $U^\prime_{m+1}(\alpha_m)$ for $m>n-l$.
\end{lemma}
\begin{proof}
Let us focus on the controlled gates (\ref{Ualpha}) with $m\ge n-l$ and consider the actions of $U(\tilde{S})$ on $\ket{0}^{\ot n}$. Since we have Eq.~(\ref{redU}), we find that the first $(n-l)$ qubits only play the r\^ole of the control qubits for Eq.~(\ref{Ualpha}). Besides, since the initial state $\ket{0}^{\ot n}$ is separable, these first $(n-l)$ qubits remain unchanged under the actions of Eq.~(\ref{Ualpha}) with $m\ge n-l$. This observation implies that the controlled gate (\ref{Ualpha}) for $\alpha_m$ becomes trivial ($\bm{1}^{\ot n}$) unless there exists an appropriate state $\ket{\alpha_{m^\prime}}\in\Bot_{i=n-l+1}^m{\cal H}_i$, by which the vector of the control qubits $\ket{\alpha_m}$ can be written as Eq.~(\ref{aa}).

If $\ket{\alpha_m}$ is written as Eq.~(\ref{aa}) and hence the first $n-l$
qubits are in the state $\ket{0}^{\ot (n-l)}$, these $n-l$ qubits play no r\^ole as control qubits. Thus, we can replace $U_{m+1}(\alpha_m)$ by $U_{m+1}^\prime(\alpha_m)$ as is claimed.
\end{proof}

\smallskip

Parallel to the previous section, we define
\bea
U^\prime_{m+1}:=\prod_{\alpha_m=0}^{2^m-1}U^\prime_{m+1}(\alpha_m)
\eea
for $m>n-l$ and
\bea
U^\prime:=U^\prime_nU^\prime_{n-1}\cdots U^\prime_{n-l+1}.
\eea
Then, from the above lemma, we immediately see
$
U^\prime\in{\cal S}(\tilde{S}).
$
Thus, hereafter we reset
\bea
U(\tilde{S})=U^\prime.
\eea
We estimate the circuit complexity for this $U(\tilde{S})$.
\begin{proposition}
The circuit complexity of $U(\tilde{S})$ has an upper bound ${\cal O}(l^22^l)$.
\end{proposition}
\begin{proof}
 Obviously there remain $m^\prime$ qubits for each possibly non-trivial 
$U_{m+1}^\prime(\alpha_m)$ as free control qubits. Since we can place either 
$\ket{0}$ or $\ket{1}$ in each qubit in $\ket{\alpha_{m^\prime}}$, the number 
of non-trivial $U_{m+1}^\prime(\alpha_m)$ is at most $2^{m^\prime}$ for 
each $m$.
Also, the number of those control qubits in each $U^\prime_{m+1}(\alpha_m)$ is manifestly $m^\prime$. Thus, utilizing the results of Ref.~17, we
estimate the circuit complexity of $U(\tilde{S})$ as
\bea
1+\sum_{m^\prime=1}^{l-1}{m^{\prime}}^22^{m^\prime}\approx l^22^l.
\eea
\end{proof}

\smallskip

\subsection{Construction of $\pi_\sigma$}

Now, we turn to the implementation of the unitary operator $\pi_\sigma$
in terms of the elementary gates. 
To this end, we first introduce
\be
B:=\{x \,|\, x\in S\backslash(S\cap\tilde{S})\}
\quad
{\rm and}
\quad
C:=\{x \,|\, x\in \tilde{S}\backslash(S\cap\tilde{S})\}.
\label{defBC}
\ee
It immediately follows from $|S|=|\tilde{S}|$ that
\be
|B|=|C|\le|S|,
\label{BCS}
\ee
which implies that we can introduce a bijection between $B$ and $C$. 
Based on this observation, we define a transposition $(x\, y)$ for any 
pair $x\in B$ and $y\in C$ . Since these transpositions commutate with 
each other, $\sigma$ can be realized as a product of transpositions for all 
distinct pairs. 

Utilizing the methodology given in Ref.~17, which is based on the 
grey code and the Hamming distance, the quantum gate of such transposition is 
implemented with at most $n$ controlled gates with $n-1$ control qubits. 
Thus, the circuit complexity of the transposition is ${\cal O}(n(n-1)^2)={\cal O}(n^3)$. Combining this observation and the inequality (\ref{BCS}) with 
the fact that we need $|B|$ transpositions to implement $\sigma$,
we conclude that
\begin{proposition}
The circuit complexity of $\pi_\sigma$ satisfying Eq.~(\ref{sigmaS}) has an upper bound ${\cal O}(n^3|S|)$.
\end{proposition}

\subsection{Circuit Complexity}

\begin{figure}[t]
(a)
\includegraphics[width=2.2in]{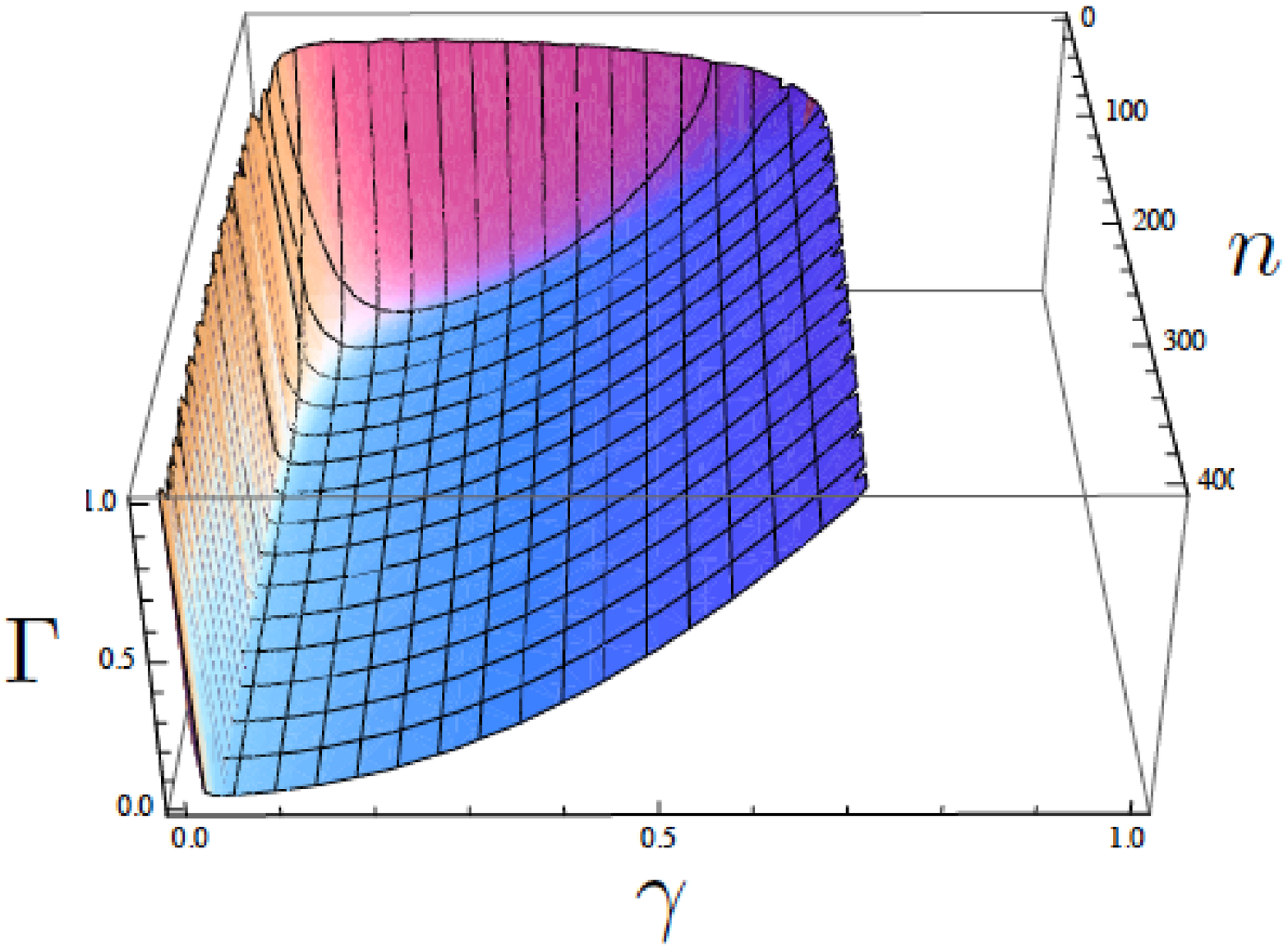}
(b)
\includegraphics[width=2.2in]{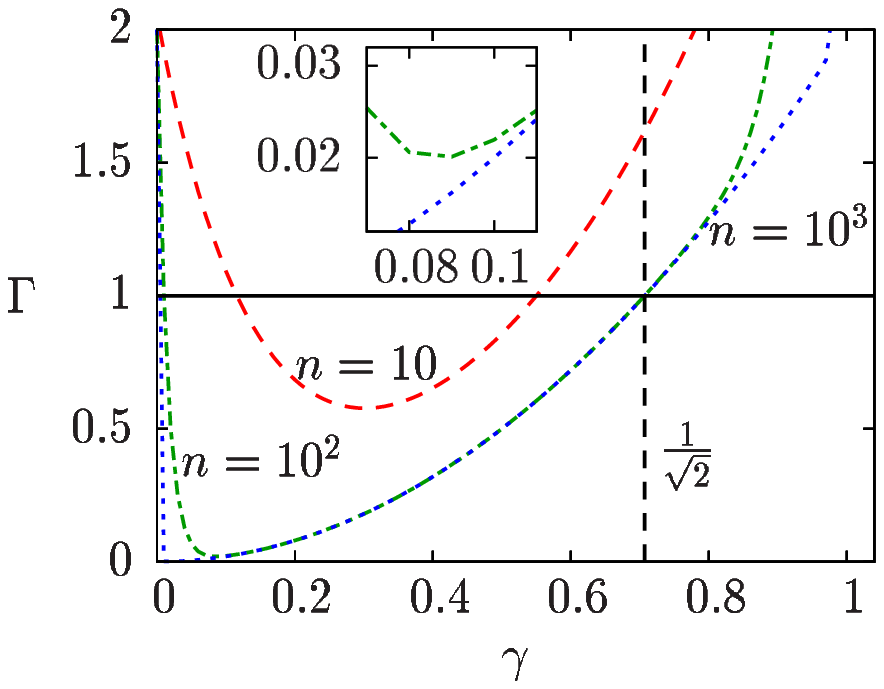}
\caption{(Color online) (a) Region of $n$ and $\gamma$ satisfying $\Gamma\le1$. Here, $\Gamma$ is defined as the ratio between the circuit complexity of Eq.~(\ref{NovelGrover}) and that of Eq.~(\ref{ConvGrover}), and it can be regarded as a
function of the number of qubits $n$ and the parameter $\gamma$ characterizing 
the size of the set of the targets $S$. (b) Graphs of $\Gamma$ with $n=10$ (dashed line), $n=100$ (dashed dotted line) and $n=1000$ (dotted line).  In the region where $\Gamma$ is less than unity, the algorithm Eq.~(\ref{NovelGrover}) dominates over the conventional algorithm Eq.~(\ref{ConvGrover}) from the viewpoint of the circuit complexity.}
\label{CCcompare}
\end{figure}

So far, we clarified the circuit complexity of each ingredient of Eq.~(\ref{NovelGrover}), which further yields
\begin{corollary}
The order of circuit complexity to obtain $\ket{\psi^{(k)}}$ by the algorithm (\ref{NovelGrover}) is
\be
{\cal O}\(2n^3|S|+2kn^2+2kl^22^l\).
\label{CCgeneral}
\ee
\end{corollary}
\begin{proof}
The first term in the argument of Eq.~(\ref{CCgeneral}) comes from Proposition 3. The second term results from the fact that the circuit complexity of $D$ and $P$ is ${\cal O}(n^2)$ and the total number of $D$ and $P$ in the r.h.s. of Eq.~(\ref{NovelGrover}) is equal to $2k$. Proposition 2 and the number of $U(\tilde{S})$ in Eq.~(\ref{NovelGrover}) yield the third term.
\end{proof}

\smallskip

Let us compare the circuit complexity given by Eq.~(\ref{CCgeneral}) with that 
for Eq.~(\ref{ConvGrover}) by setting $k\approx\sqrt{|N|/|S|}$. 
For this purpose, it is convenient to introduce the ratio
\bea
\Gamma:=\frac{2n^3|S|+2\(n^2+l^22^l\)\sqrt{|N|/|S|}}{n^2(|S|+1)\sqrt{|N|/|S|}},
\eea
where the numerator comes from Eq.~(\ref{CCgeneral}) and the denominator comes from the circuit complexity ${\cal O}(n^2|S|)$ of $O_{\rm conv}(S)$. Note that the circuit complexity of $O_{\rm conv}(S)$ cannot be improved even if we employ 
the method introduce in this section (See Appendix).  From the viewpoint of the circuit complexity, we conclude that Eq.~(\ref{NovelGrover}) dominates
when $\Gamma<1$. Recalling $|N|=2^n$, $|S|=|\tilde{S}|<2^l$ and introducing a 
ratio $\gamma:=l/n$, the ratio $\Gamma$ is approximated as
\bea
\Gamma\approx 2\[n2^{-\frac{n}{2}(1-\gamma)}+(2^{-n\gamma}+\gamma^2)\].
\eea
If $\gamma$ is fixed and we take the limit $n\rightarrow\infty$, we obtain $\Gamma\rightarrow2\gamma^2$ from which we find that the algorithm (\ref{NovelGrover}) is preferable if $\gamma$ satisfies
\bea
\gamma<\frac{1}{\sqrt{2}}\approx0.71.
\eea
Further, Fig.~\ref{CCcompare} tells us that the above bound $\gamma=1/\sqrt{2}$
is universal if the number of qubits is sufficiently large. Thus, we may 
conclude for a sufficiently large database that the proposed algorithm is 
preferable than the conventional one from the viewpoint of the 
circuit complexity,
provided that the number of the target satisfies
\bea
|S|\lesssim |N|^{0.71}.
\eea

\section{Examples}
In this section, we give two simple examples
to demonstrate the difference between the implementations of
$U(S)$ and $U(\tilde{S})$. For brevity, we utilize the binary representations 
to describe the elements of sets $S$, $\tilde{S}$ and the sets derived from 
them. For the later convenience, we also introduce a one-qubit gate
\bea
V:=\frac{1}{\sqrt{2}}\(\bm{1}-\im\sigma_y\).
\eea
Note that
$
V\ket{0}=H\ket{0},
$
where $H$ is the Hadamard gate
\be
H:=\frac{1}{\sqrt{2}}\(\sigma_x+\sigma_z\).
\label{Hadamard}
\ee

\subsection{Circuit for $U(S)$}

As a demonstration of our algorithm to design the relevant
quantum circuit, let us consider a database search problem to extract
four three-bit strings
\be
S=\{000, 001, 010,100\}.
\label{exS}
\ee 
The classification of whose elements according to the dichotomy yields
\bea
S_1(0)=\{000, 001, 010\},
\qquad
S_1(1)=\{100\},
\eea
from which we find
\be
V_1=\frac{1}{2}\(\sqrt{3}\times\bm{1}-\im\sigma_y\).
\label{V1S}
\ee
Further, from the dichotomy on the elements in $S_1(0)$ and $S_1(1)$, we obtain
\begin{eqnarray*}
S_2(0,0)&=&\{000, 001\}
\qquad
S_2(1,0)=\{010\},\\
S_2(0,1)&=&\{100\},
\qquad\quad\,\,\,
S_2(1,1)=\emptyset,
\end{eqnarray*}
respectively. These enable us to construct
\be
V_2(0)=\frac{1}{\sqrt{3}}\(\sqrt{2}\times\bm{1}-\im\sigma_y\)
\quad
{\rm and}
\quad
V_2(1)=\bm{1}.
\label{V2S}
\ee
\begin{figure}[t]
\begin{center}
\includegraphics[width=3in]{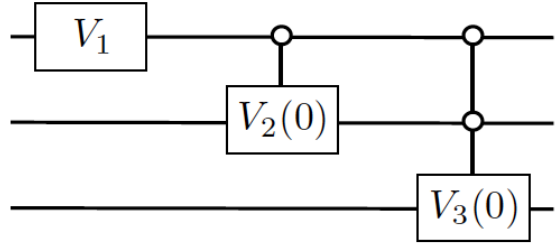}
\caption{Quantum circuit of $U(S)$ for $S$ given in Eq.~(\ref{exS}). The explicit form of the one-qubit gates $V_1$, $V_2(0)$ and $V_3(0)$ are given in Eqs.~(\ref{V1S}), (\ref{V2S}) and (\ref{V3S}), respectively.}
\label{U01}
\end{center}
\end{figure}
By the same way, we find
\begin{eqnarray*}
&&S_3(0,0)=\{000\},
\qquad
S_3(1,0)=\{001\},\\
&&S_3(0,1)=\{010\},
\qquad
S_3(1,1)=\emptyset,\\
&&S_3(0,2)=\{100\},
\qquad
S_3(1,2)=\emptyset,\\
&&S_3(0,3)=S_3(1,3)=\emptyset.
\end{eqnarray*}
Based on these sets, we have
\be
V_3(0)=V,
\qquad
V_3(1)=V_3(2)=V_3(3)=\bm{1}.
\label{V3S}
\ee
These results are summarized as the quantum circuit depicted in Fig. \ref{U01}.

\subsection{Circuit for $U(\tilde{S})$}

\begin{figure}[t]
\begin{center}
\includegraphics[width=3in]{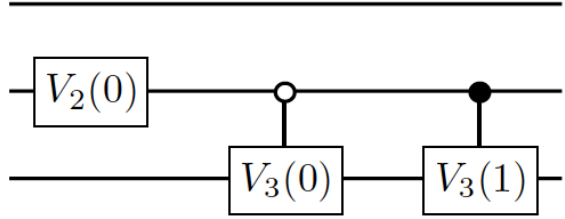}
\caption{Quantum circuit of $U(\tilde{S})$ for $\tilde{S}$ given in Eq.~(\ref{exStilde}). The explicit form of the one-qubit gates $V_2(0)$, $V_3(0)$ and $V_3(1)$ are given in Eqs.~(\ref{V1t}) and (\ref{V2t}). Merging the second and third control gates, this circuit turns into a gate composed from two one-qubit gates.}
\label{U01tilde}
\end{center}
\end{figure}

From the given set $S$ in Eq.~(\ref{exS}), we construct the set $\tilde{S}$ as
\be
\tilde{S}=\{000, 001, 010, 011\}.
\label{exStilde}
\ee
By the same method utilized in the previous subsection, we find
\be
V_1=\bm{1}
\label{V1t}
\ee
as is expected from Eq.~(\ref{redU}). Also, further dichotomy reveals
\begin{eqnarray}
&V_2(0)=V_3(0)=V_3(1)=V,\\
&V_2(1)=V_3(2)=V_3(3)=\bm{1}.
\label{V2t}
\end{eqnarray}
Thus, the non-trivial $U_{m+1}(\alpha_m)$ are found from Eq.~(\ref{Ualpha})
as
\begin{eqnarray*}
U_2(0)&=&\(\ket{0}\bra{0}\ot V+\ket{1}\bra{1}\ot\bm{1}\)\ot\bm{1},\\
U_3(0)&=&\ket{00}\bra{00}\ot V+(\bm{1}^{\ot2}-\ket{00}\bra{00})\ot\bm{1},\\
U_3(1)&=&\ket{01}\bra{01}\ot V+(\bm{1}^{\ot2}-\ket{01}\bra{01})\ot\bm{1},\\
\end{eqnarray*}
by the help of the resolution of identity. On the other hand, 
following Eqs.~(\ref{Uprimealpha1}), (\ref{Uprimealpha2}) and (\ref{Uprimealpha3}), we introduce
\begin{eqnarray*}
U_2^\prime(0)&=&\bm{1}\ot V\ot\bm{1},\\
U_3^\prime(0)&=&\bm{1}\ot\(\ket{0}\bra{0}\ot V+\ket{1}\bra{1}\ot\bm{1}\),\\
U_3^\prime(1)&=&\bm{1}\ot\(\ket{1}\bra{1}\ot V+\ket{0}\bra{0}\ot\bm{1}\).\\
\end{eqnarray*}
Then one can verify, by direct calculation, that
\bea
U_3(1)U_3(0)U_2(0)\ket{0}^{\ot3}=U_3^\prime(1)U_3^\prime(0)U_2^\prime(0)\ket{0}^{\ot3},
\eea
which is the key for the reduction of the circuit complexity. The quantum circuit obtained by this replacement is shown in Fig.~\ref{U01tilde}. 
We further observe that
\bea
U_3^\prime(1)U_3^\prime(0)U_2^\prime(0)\ket{0}^{\ot3}=\bm{1}\ot V\ot V\ket{0}^{\ot 3}=\bm{1}\ot H\ot H\ket{0}^{\ot 3},
\eea
which implies that our algorithm yields an intuitive way to produce $\ket{\tilde{S}}$.

\subsection{Circuit for $\pi_\sigma$}
\begin{figure}[t]
\begin{center}
\includegraphics[width=3in]{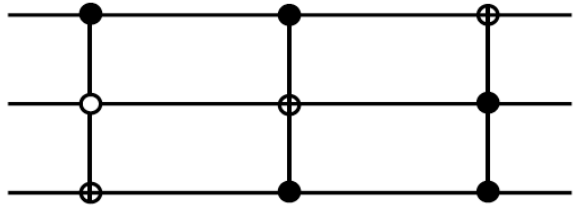}
\caption{Quantum circuit of $\pi_\sigma$ for the permutation (\ref{decSigma}).}
\label{pi_sigma}
\end{center}
\end{figure}

For $S$ and $\tilde{S}$ in this section, we find
\bea
B=\{100\}
\qquad
{\rm and}
\qquad
C=\{011\}
\eea
from Eq.~(\ref{defBC}). Then, the permutation $\sigma$ associated with $B$ and $C$ is uniquely determined,
and we find
\bea
\sigma=(100\,\, 011).
\eea
We design a sequence of bit strings from $100$ to $011$ so that the Hamming
distance between the neighbours is equal to the identity, {\it e.g.},
\bea
100, 101,111,011.
\eea
This sequence shows that the permutation operator $\pi_\sigma$ is rewritten
as
\be
\sigma=(100\,\, 101)(101\,\, 111)(111\,\, 011).
\label{decSigma}
\ee
Since every transposition in the RHS of Eq.~(\ref{decSigma}) is realized
by a controlled-controlled-NOT gate, we can materialize $\pi_{\sigma}$ 
as a quantum circuit of these controlled-controlled-NOT gates (see Fig.~\ref{pi_sigma}).

\section{Conclusion and Discussions}
In this paper, we proposed a systematic construction and implementation
of operators 
suitable for demonstration of quantum database search algorithm 
in the sense of Ref.~2 with multiple targets. 
The operator $O(S)$ is introduced for this purpose and it 
is proven that it works as an oracle for GA under a
certain condition on the initial state (Theorem 1). Since the modified 
oracle $O(S)$ is decomposed into operators $P$ and $U(S)$, where $P$
is the conventional oracle operator with a unique target
$\ket{0}^{\ot n}$, we have shown 
that the quantum circuit of the oracle built by the elementary gate set can 
be designed if that of $U(S)$ was given.  
The construction of the quantum circuit
for $U(S)$ was accomplished by employing Lemma 2 and Theorem 2, 
which were derived through dichotomy on the value of each qubit.
We should emphasize that our work takes advantance of 
the non-uniqueness of the implementation of a quantum information processing. 

Also, we found another algorithm to perform the same database search (Theorem 3). One of the advantages of utilizing this is that we can reduce the circuit 
complexity considerably (Proposition 2). We showed that there existed a simple 
condition under which our algorithm was more advantageous than the conventional algorithm with the oracle $O_{\rm conv}(S)$ from the viewpoint of the circuit 
complexity.

The modified oracle $O(S)$ fails to sort the targets with a good precision 
when noise and errors are present. This deviation from the targets will be 
circumvented to some extent by fault tolerant quantum error correction codes 
(QECC)$^{20}$, as in the case of the conventional sorting algorithm with $O_{\rm conv}(S)$.
QECC can be built-in in $O(S)$ similarly to $O_{\rm conv}(S)$, since QECCs are 
independent of the system size and actual forms of quantum algorithms.
Note, however, that the algorithm is probabilistic in nature. We have to verify the targets obtained to make sure that we got correct results.

\section*{Acknowledgments}

We would like to thank Akira SaiToh for valuable discussions. This work is partially supported by \lq Open Research Center\rq~Project for Private Universities: matching fund subsidy from MEXT, Japan. 

\section*{Appendix.}
In this Appendix, we decompose the operator $O_{\rm conv}(S)$ into 
the elementary gates. To this end, we note that $\braket{x|y}=\delta_{xy}$.
Due to this orthogonality, we have
\begin{eqnarray}
O_{\rm conv}(S)&=\bm{1}^{\ot n}-2\sum_{x\in S}\ket{x}\bra{x}
=\prod_{x\in S}\(\bm{1}^{\ot n}-2\ket{x}\bra{x}\)
=\prod_{x\in S}U_xPU_x^\dag,
\label{decOconv}
\end{eqnarray}
where $U_x$ is a unitary operator satisfying
\be
U_x\ket{0}^{\ot n}=\ket{x}.
\label{Ux}
\ee
It is obvious that $U_x$ is far from unique. Nonetheless, in view of 
Eq.~(\ref{xstate}), 
there exist local unitary operators satisfying Eq.~(\ref{Ux}), {\it e.g.},
\be
U_x=\Bot_{i=0}^{n-1}\(\bm{1}\delta_{x_i0}+\sigma_x\delta_{x_i1}\).
\label{Uxone}
\ee
We estimate the circuit complexity of $O_{\rm conv}(S)$ using Eq.~(\ref{Uxone}).
Since the number of (non-trivial) one-qubit gates to construct $U_x$ is bounded
by $\sum_ix_i\le n$ from above, plugging this into the number of gates for $P$, we conclude that we need ${\cal O}(n^2|S|)$ elementary gates to
implement $O_{\rm conv}(S)$.

Let us finally comment on the circuit complexity of $O_{\rm conv}(\tilde{S})$, 
where $\tilde{S}$ is given by Eq.~(\ref{sigmaS}). Since we have 
$|S|=|\tilde{S}|$ and the upper bound of the circuit complexity of $U_x$ is 
$n$, the circuit complexity is left {\it unchanged} even if we employ the 
decomposition (\ref{decOconv}) for $\tilde{S}$. This observation implies that 
the circuit complexity cannot be improved by the method proposed in Sec.~IV
as far as the decomposition (\ref{decOconv}) for $O_{\rm conv}(S)$ is utilized.

\vspace*{-6pt}   

\section*{References}


\vspace*{-5pt}   

\myitem
{P. W. Shor, in
{\it Proc. 35th Annual Sympo. on Foundations
of Computer Science} (IEEE Computer Society
Press, 1994) pp.~124.}

\myitem
{L. K. Grover,
in {\it Proc. 28th Annual ACM Sympo. on the Theory of Computing} (1996), pp.~212.
}

%
\myitem
{L. K. Grover, 
\prl{79}{1997}{325}.}

\myitem
{M. Boyer, G. Brassard, P. H\o yer and A. Tapp,
\fp{46}{1998}{493}.}

\myitem
{M. A. Nielsen and I. L. Chuang,  
{\it Quantum Computation and Quantum Information}
(Cambridge University Press, 2000).}

\myitem
{M. Nakahara and T. Ohmi,
{\it Quantum Computing: From Linear Algebra to Physical Realizations}
(Boca Raton: Taylor and Francis, 2008).}

\myitem
{A. Younes, 
Strength and Weakness in Grover's Quantum Search Algorithm, 
arXiv:0811.4481.}

\myitem
{I. L. Chuang, N. Gershenfeld and M. Kubinec,
\prl{80}{1998}{3408}.}

\myitem
{L. M. K. Vandersypen, M. Steffen, M. H. Sherwood, C. S. Yannoni, G. Breyta and I. L. Chuang,
{\it Appl. Phys. Lett.} {\bf 76} (2000) 646.}

\myitem
{Z. Diao, M. S. Zubairy and G. Chen,
{\it Z. Naturforsch.} {\bf 57a} (2002) 701.}

\myitem
{V. L. Ermakov and B. M. Fung,
\prA{66}{2002}{042310}.}

\myitem
{M. S. Anwar, D.  Blazina, H. A.  Carteret, S. B. Duckett and J. A. Jones,
{\it Chem. Phys. Lett.} {\bf 400} (2004) 94.}

%
\myitem
{P. Walther, K. J. Resch, T. Rudolph, E. Schenck, H. Weinfurter, V. Vedral, M. Aspelmeyer and A. Zeilinger,
{\it Nature} {\bf 434} (2005) 169.}

%
\myitem
{K.-A. Brickman, P. C. Haljan, P. J. Lee, M. Acton, L. Deslauriers and C. Monroe,
\prA{72}{2005}{050306(R)}.}

\myitem
{N. D. Mermin,
{\it Quantum Computer Science}
(Cambridge University Press, 2007).}

\myitem
{I. E. Linington, P. A. Ivanov and N. V. Vitanov,
\prA{79}{2009}{012322}.}

\myitem
{A. Barenco, C. H. Bennett, R. Cleve, D. P. DiVincenzo, N. Margolus, P. Shor, T. Sleator, J. A. Smolin and H. Weinfurter,
\prA{52}{1995}{3457}.}

\myitem
{M. M\"ott\"onen, J. J. Vartiainen, V. Bergholm and M. M. Salomaa,
\prl{93}{2004}{130502}.}

\myitem
{V. Bergholm, J. J. Vartiainen, M. M\"ott\"onen and M. M. Salomaa,
\prA{71}{2005}{052330}.}

\myitem
{F. Gaitan,
{\it Quantum Error Correction and Fault Tolerant Quantum Computing},
(Boca Raton: Taylor and Francis, 2008).}
\end{document}